\documentclass[aps,preprint,epsfig,multicol,nofootinbib,showpacs]{revtex4}
\usepackage{graphicx}
\usepackage{amsmath, amsthm, amssymb}
\usepackage{comment}
\usepackage{color}
\begin{document}
\title{Epidemics on Interconnected Networks}
\author{M. Dickison,$^{1}$ S. Havlin,$^{1,2}$ and H. E. Stanley$^{1}$ }
\affiliation{$^1$Center for Polymer Studies, Physics Department, Boston University, Boston, Massachusetts 02215, USA.\\
$^2$Minerva Center, Department of Physics, Bar Illan University, Ramat Gan, Israel.}
\date{\today}

\begin{abstract}
Populations are seldom completely isolated from their environment. Individuals in a particular geographic or social region may be considered a distinct network due to strong local ties, but will also interact with individuals in other networks. We study the susceptible-infected-recovered (SIR) process on interconnected network systems, and find two distinct regimes.  In strongly-coupled network systems, epidemics occur simultaneously across the entire system at a critical infection strength $\beta_c$, below which the disease does not spread.  In contrast, in weakly-coupled network systems, a mixed phase exists below $\beta_c$ of the coupled network system, where an epidemic occurs in one network but does not spread to the coupled network.  We derive an expression for the network and disease parameters that allow this mixed phase and verify it numerically. Public health implications of communities comprising these two classes of network systems are also mentioned.
\end{abstract}
\maketitle{}

\section{Introduction}
\begin{figure}
\includegraphics[width=8cm,height=5cm]{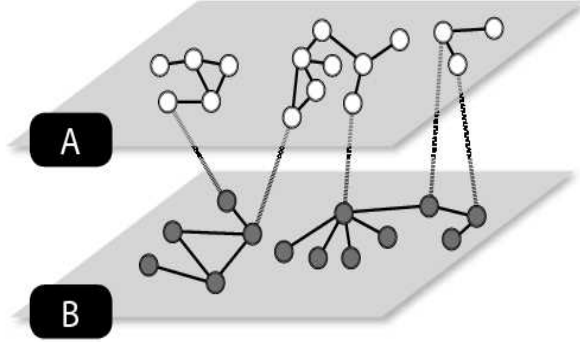}
\caption{An interconnected network system with two networks: A and B.  Nodes have intranetwork links within their own network, but also internetwork links connecting them to the other network.
\label{interconnecting}}
\end{figure}

Complex network models of the interactions in human society have been used to
understand many problems in epidemiology~\cite{Pastor,Kuperman,Newman,Cohen2003,Vespignani-book,Parshani,Lagorio,Morris_book}. These models have generally assumed
that all of the nodes interact on a single network with a single degree
distribution. Even when these degree distributions allow for large
heterogeneities---as in the case of scale-free networks~\cite{Barabasi-Nature}, where hubs with large
numbers of connections can arise---the assumption remains that every node is part
of a single network and is represented by a single underlying topology. In reality,
however, societies are composed of many interconnected networks, as in Fig.~\ref{interconnecting}, which may be
communities within a larger population or separate systems entirely. A disease
can spread through the network of direct personal contacts, via the water
utilities network, and through travel from city to city over highway or airline
networks.  These interconnected network systems may be comprised of different types of
nodes, which may have degrees drawn from distinct degree distributions, and may
have different connectivities between them. {Real world examples of these systems can be
seen in a 2009 study by Stehl\`e et. al. which found a three fold difference in interaction 
time between students inside and outside of their own class~\cite{Stehle}. Other studies have shown similar patterns~\cite{Salathe,Cattuto}. Human-animal interacting network systems are also of great importance. The H5N1 variant of influenza spreads through the network of birds, and from them to individuals in the network of humans that work or live
closely with them. While no current mechanism exists for efficient spreading from human to human,
there is substantial concern that such a mechanism will evolve~\cite{Liu}.  Human-mosquito-human
transmission for the {\it P. knowlesi} malaria strain is also a worrying concern~\cite{Kantele}.}

Interconnected network systems have been of interest to researchers in
numerous different ways~\cite{Kurant,Allard,Dorogovtsev,Ott}. Interconnected
dependency networks, where failure in nodes in one network causes failures of
dependent nodes in the other network, exhibit failure cascades, where the
cross network dependencies result in a network much more easily fragmented than
single networks of the same degree distribution~\cite{Buldyrev2010}. Interconnected power networks, where transport
capacity and failure vulnerability are competing properties, were examined and
an optimal level of interconnection found~\cite{D'Souza}. Networks without
dependencies, such as interconnected social networks, where populations exist at
city, state, and national levels, have also been examined. In these networks, the level of
movement between cities (the interconnections between them) have been shown to
affect the epidemic transition on the metapopulation level
\cite{Colizza,Colizza_metapop}, although in this case the low-level networks
were treated in a mean field fashion, classified only by rate equations and
infection numbers, with no internal network features. In addition, the
percolation threshold in interacting networks was found to be lower than in
single networks, with a giant cluster appearing for a smaller total number of
links~\cite{Leicht2009}.

In this work we consider two interconnected networks (or, alternately,
interconnected communities within a single, larger network). We pose the
question: Under what conditions will an epidemic spread only on the sub-networks,
with minimal isolated infections on other network components, and under what
conditions will it spread across the entire interconnected network system?
Depending on the parameters of the individual networks and their
interconnections, connecting one network to another can have a profound or a small
effect on the spread of an epidemic. Identifying the conditions in which these
cases occur is vital to our understanding and management of epidemic processes.

We define two different interconnected network regimes, strongly and weakly coupled, and find the
interaction strength value separating these two regimes. Our primary
result is to show that in the strongly-coupled case, we find that all networks
are simultaneously either disease free or part of an epidemic, while in the
weakly-coupled case a new ``mixed" phase can exist. In this mixed phase, the
disease is epidemic on only one network, and not in other networks, despite the interconnections.
 The applications to public health are straightforward. If two neighboring
communities comprise a strongly-coupled network system, then an outbreak in any
community is cause for immediate concern in the other. Due to this, in the
strongly-coupled case it becomes important to pursue a strategy of
communication and joint action between public health agencies, and perhaps even intervention from a single
agency with higher authority.

\section{Model}
In this section, we consider the case of only two interconnected networks of equal size,
but it is easily possible to extend the model to an arbitrary number of networks of any size.

We form our interconnected network systems in the following way:
\begin{enumerate}
\item{Generate two networks, A and B, with their own intranetwork (A
$\leftrightarrow$ A and B $\leftrightarrow$ B) degree distributions $P_A(k)$ and
$P_B(k)$ according to the standard Molloy-Reed configuration
model~\cite{Molloy-Reed}}.
\item{Draw a degree from the internetwork (A $\leftrightarrow$ B) degree
distribution, $P_{AB}(k)$, for all the nodes in both networks.}
\item{If the total degree assigned to nodes in network A is not equal to the
total degree assigned to nodes in network B, randomly reassign a node in
B until the total numbers in each network are equal.}
\item{Randomly connect nodes in network A to nodes in network B to form the
interconnected network system.}
\end{enumerate}
This method generates random, uncorrelated, interconnected network systems with specified
inter- and intra-network degree distributions. While this method works for any
arbitrary degree distribution, $P_x(k)$, we present results only for random
Poissonian degree distributions.

The susceptible-infected-recovered (SIR) epidemic model is used here to study
the effects of interconnected network structure on epidemic threshold. The SIR
model is well established and describes diseases such as HPV, seasonal
influenza, or H1N1~\cite{Colizza,Anderson_May}. In this model, each node has
three possible states: susceptible ($s$), infected ($i$), or recovered ($r$).
Each node begins in state $s$, except for a single node in one network chosen to be in state $i$.
Nodes in state $i$ infect their neighbors in state $s$ with probability $\beta$
at each time step, changing them to $i$. Nodes enter state $r$ after spending a
recovery time $t_r$ in state $i$.

In order to find the threshold for an epidemic, we can think of epidemic
spreading as a bond percolation process~\cite{Newman,Bunde_book,Havlin_book} on a
network. In bond percolation, links between nodes are activated with a certain
probability $p$. If this probability is greater than a certain critical value,
$p_c$, then a giant cluster emerges, where the existence of a path between any
two nodes is almost certain. In a disease-spreading model, nodes infect their
neighbors, ``activating" the links between them with a certain probability, and a
disease reaches nodes through this entire network above a certain critical
value, $\beta_c$, just as in the case for percolation.

In complex networks, this critical threshold for percolation if all potential links are
activated is $\kappa=2$. Here $\kappa$ is the expected number of nearest
neighbors that a node chosen by following an arbitrary link will have, and is
calculated from the ratio between the second and the first moments of the degree distribution:
$\kappa = \langle k^2 \rangle / \langle k \rangle$. For $\kappa \geq 2$, a giant
cluster exists, while for $\kappa \leq 2$ only small isolated clusters exist. If
some subset of bonds is activated at random with probability $p$, a giant
cluster appears at a critical value of $p_c=1/(\kappa -1)$~\cite{Cohen}.

The SIR model likewise has an epidemic phase transition at a critical
$\beta=\beta_c$ below which the disease remains confined to the local
neighborhood of the initial infection, and above which the disease spreads
throughout the network. This transition from the disease-free
phase to the epidemic phase depends on the average number of secondary
infections per infected node becoming larger than one. This allows the long-term
survival of the disease, as the infection density will grow over time on
average, and thus ensure that the epidemic spreads to a large
fraction of the population. In our problem, the expected number of susceptible
neighbors that a node has when it just becomes infected is given by $\kappa -1$,
since the total expected number of neighbors is $\kappa$, and one of them must
be excluded as the infected parent from which the current node descended. The
transmissiblity $T_{\beta}= 1-(1-\beta)^{t_r}$ is the probability to infect a neighbor before recovery. The mean number of secondary
infections per infected node is thus
$N_I = (\kappa -1) T_{\beta}.$
The infection will die out if each infected node does not infect on average at
least one
replacement  so, for a very large network, the critical point is given by the
relation $(\kappa -1) T_{\beta}=1$.
The single network model exhibits only a single transition at $\beta_c$ given by
\cite{Newman}
\begin{equation}
\beta_c(\kappa) =1- \left[1-(\kappa-1)^{-1}\right]^{1/t_r}.
\label{transition}
\end{equation}
In the interconnected network model, the behavior is more complicated, as the
disease can potentially cause an epidemic in different combinations of the networks.
The disease can either be in the epidemic phase in both networks, in the disease-free
phase in both networks, or active in one network while the other remains
disease free, called here the mixed phase. The boundaries of these phases are
controlled by $\kappa_A$, $\kappa_B$, and $\kappa_T$, where $\kappa_A$ and
$\kappa_B$ are calculated over the individual A and B networks, disregarding
internetwork connections, and $\kappa_T$ is calculated over the entire coupled network system,
including intra- and inter- network links.

\section{Strongly-Coupled Network Systems}
\begin{figure}
\includegraphics[width=6cm,height=8cm, angle=-90]{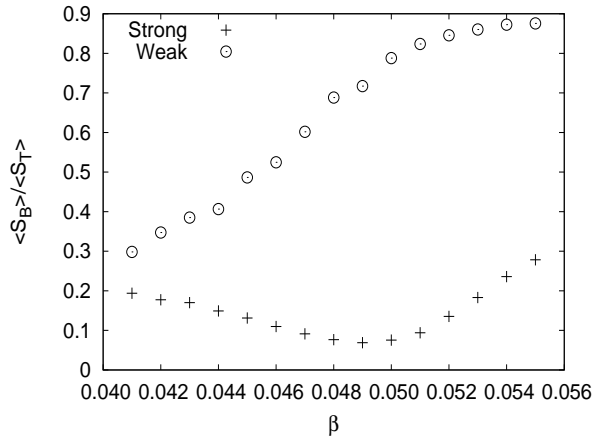}
\caption{Epidemics in strongly-coupled network systems spread across all networks,
remaining confined to one network in weakly-coupled network systems. We plot $S_B$, the size
of the largest connected cluster solely in network B, divided by $S_T$, the size of the
infected cluster across the interconnected network system, for both strongly- and weakly-coupled network systems. The B only cluster decreases in relative size until criticality ($\beta_c(\kappa_B)=.048$), showing
that the epidemic spreads throughout both networks rather than remaining confined in B in the strongly-coupled system.
 By contrast, in the weakly-coupled system, the relative size of the B only cluster grows until
$\beta=\beta_c(\kappa_T)=.054$, showing that growth is localized in the more
strongly coupled network. For the strongly-coupled network, $\langle k_A
\rangle =1.5,\langle k_B \rangle = 2.5$, and $\langle k_{AB} \rangle = 2.5$. For
the weakly-coupled network, $\langle k_A \rangle =1.5,\langle k_B \rangle =
4.55$, and $\langle k_{AB} \rangle = 0.3$. In all cases, $N_A=N_B=10^4$ and $t_r=5$.}
\label{cluster}
\end{figure}

We consider an interconnected network system to be strongly coupled if $\kappa_T$ is
larger than $\kappa_A$, and $\kappa_B$.
{For random networks, $\kappa = \langle k \rangle +1$, and thus we may write $\kappa_T$
in terms of the average degrees $\langle k_A \rangle$, $\langle k_B \rangle$, and $\langle k_{AB} \rangle$
as follows:
\begin{eqnarray}
\nonumber \kappa_T &=& [\langle k_A \rangle^2 + \langle k_A \rangle +\langle k_B \rangle^2 + \langle k_B \rangle    + 2\langle k_{AB}\rangle^2+2\langle k_{AB} \rangle \\&& \nonumber + 2\langle k_A \rangle \langle k_{AB} \rangle +  2\langle k_B \rangle \langle k_{AB}\rangle ]\\&&[ \langle k_A \rangle + \langle k_B \rangle + 2\langle k_{AB} \rangle ]^{-1}.
\label{kappa_T}
\end{eqnarray}
Without loss of
generality, we define network $B$ as the more intraconnected network ($\langle k_B \rangle > \langle k_A \rangle$).
For fixed network parameters $\langle k_A \rangle$ and $\langle k_B \rangle$, we can then derive the critical interaction
strength $\langle k_{AB}\rangle_c$ that separates strongly-coupled ($\kappa_T > \kappa_B$) from weakly-coupled ($\kappa_T < \kappa_B$) networks:
\begin{equation}
\langle k_{AB}\rangle_c =  \frac{\sqrt{2\langle k_A\rangle \langle k_B\rangle - \langle k_A \rangle ^2}-\langle k_A \rangle}{2}.
\label{threshold}
\end{equation}}
In strongly-coupled network systems, we
expect any epidemic to emerge simultaneously on networks A and B. Using Eq.
\ref{transition} for each of the three $\kappa$, it can be seen that for the
strongly-coupled case $\beta_c(\kappa_T)$, the critical value of $\beta$ for
the disease to emerge on the giant component formed by the entire interconnected
network, is smaller than both $\beta_c(\kappa_A)$ and $\beta_c(\kappa_B)$, the
critical values of $\beta$ for epidemics to spread on networks A or B ignoring
internetwork links. As such, any pathogen virulent enough to spread in network A
or B alone will have already caused an epidemic occurring across the
interconnected network system. For this case, the disease spreads across the
interconnected network system as a single network, with the internetwork connections
bringing an epidemic into existence before any intranetwork connections can do
so independently; the mixed phase will not be seen.  
To support this, we plot
the ratio of the largest connected infected cluster formed solely from nodes
connected with intranetwork links, compared to the size of the largest
connected cluster formed by nodes connected with all links, in Fig.
\ref{cluster}. For a strongly-coupled network system, the relative size of the
largest connected infected component contained entirely in a single network
decreases initially, showing that the epidemic is occurring across the
interconnected network system, not locally in one of the networks. Thus in the strongly-coupled
case, epidemic spreading is enhanced due to internetwork
connections, with epidemics occurring for less virulent diseases than would spread on either network alone (lower $\beta_c$.)

{One note is that for networks of identical intranetwork degree ($\langle k_A \rangle = \langle k_B \rangle$)
$\langle k_{AB} \rangle_c = 0$.  That is to say, identical networks always form strongly-coupled network systems.
This is in agreement with our findings that the phase diagram of strongly-coupled network systems is similar to that
of single networks. An interacting network system formed by attaching the labels 'A' and 'B' to different halves of
a single network would be such an example system, and one should not expect that this relabeling could have any effect on the physical
properties, such as phase transitions, of that network.}

\section{Weakly-Coupled Network Systems}
{If two networks are connected with $\langle k_{AB}\rangle$ below the threshold value from Eq.~\ref{threshold}, i.e. $\kappa_B > \kappa_T$,
we define the interconnected network system to
be weakly coupled. From Eq.~\ref{kappa_T} this also gives $\kappa_T > \kappa_A$.
Turning again to Eq.~\ref{transition} we see that $\beta_c(\kappa_B) < \beta_c(\kappa_T) < \beta_c(\kappa_A)$. }

Epidemic spreading is a non-competitive process. Adding more links to a network
can only increase the spread of an epidemic, never decrease it, as the chance
of a node infecting its neighbors is constant regardless of degree. 
Thus, a disease with $\beta$ above the individual
epidemic threshold of network B ($\beta_c(\kappa_B))$ will enter the epidemic
phase on that network, regardless of the other network and the values of
$\kappa_T$ and $\kappa_A$. If $\beta$ is below $\beta_c(\kappa_T)$, however, the
disease cannot spread to more than isolated small clusters of network A. Thus, in the weakly-coupled case, we expect to see a mixed phase, with the boundaries dependent
on the values of $\beta$ and $\langle k_{AB} \rangle$. A mixed phase indicates
that the addition of the interconnections between the two networks is only affecting
epidemic spreading on the network with weaker intranetwork connections, with the
epidemic on the network with stronger intranetwork connections unchanged by the
internetwork links. The weakly-coupled
case in Fig.~\ref{cluster} shows this, with the largest connected
cluster contained entirely in B becoming larger compared to the size of the giant component with increasing
$\beta$ until $\beta_c(\kappa_T)$ is reached, indicating that the disease does not spread through connected regions
of network A.

\begin{figure}
\includegraphics[width=6cm,height=8cm, angle=-90]{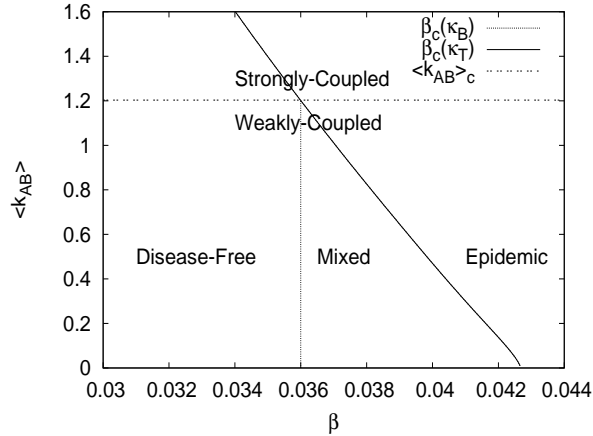}
\caption{{The mixed phase disappears in the transition from weakly- to strongly-coupled network systems.
Sample phase diagram for interacting network systems with $\langle k_A \rangle = 1.5$ and $\langle k_B
\rangle = 6.0$ as a function of infection strength $\beta$ and internetwork
degree $\langle k_{AB} \rangle$, showing the two critical $\beta$ and $\langle k_{AB} \rangle_c$.
In the weakly coupled case, below $\beta_c(\kappa_B)$ no epidemic
occurs. For $\beta_c(\kappa_B) < \beta < \beta_c(\kappa_T)$, there exists a mixed phase,
where an finite fraction of network B becomes infected, but network A has
only small infected clusters. Above $\beta_c(\kappa_T)$, an epidemic
occurs across the entire network in both the weakly- and strongly-coupled cases.}
For this diagram, $N_A=N_B=10^4$ and $t_r=5$. \label{Phase} }
\end{figure}

If $\beta$ is increased to above $\beta_c(\kappa_T)$, network B becomes capable
of spreading the disease to network A, which now enters the epidemic phase, even
for $\beta < \beta_c(\kappa_A)$. This matches the work done by Leicht and D'Souza~\cite{D'Souza} where a giant cluster
forms consisting of nodes in both networks, even when the less intraconnected network
is below its own percolation threshold.
We plot the full phase diagram for both weakly- and strongly-coupled networks  in Fig. \ref{Phase},
showing the disease-free phase, the mixed phase, the epidemic phase, and the transition between weakly- and
strongly coupled networks.
The existence of this mixed phase is important in the real-world context of
interacting networks, as the communities or systems that comprise the components
are likely to be governed by different bodies. If two cities, for example,
together form a weakly-coupled network system, the more highly connected city can
more safely disregard the links to, and response of, the less highly connected
city, as the spread of the epidemic will depend on local parameters only.

\begin{figure}
\includegraphics[width=6cm,height=8cm, angle=-90]{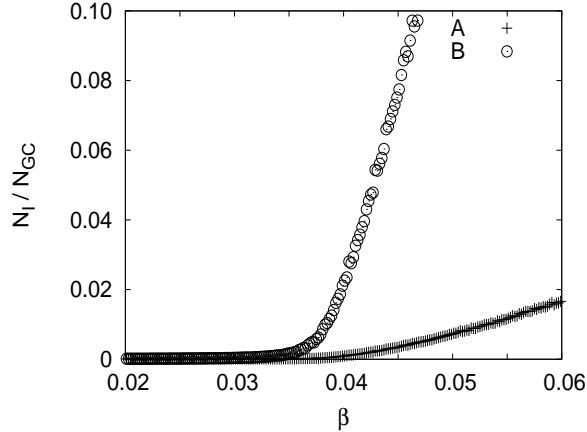}
\caption{In the mixed phase, the two networks have separate transition values.
Ratio of total number of infected $N_I$ in each network to the size of the giant cluster $N_{GC}$
for two weakly coupled networks with $\langle
k_A\rangle=1.5$, $\langle k_B\rangle=6.0$ and $\langle k_{AB}\rangle=.1$. The
respective epidemic thresholds calculated from Eq. (\ref{transition}) are
$\beta_c(\kappa_B) \approx 0.035$, $\beta_c(\kappa_T) \approx 0.0425$. The
infection can be seen to become epidemic in network B well before it does in
network A. 
The network of networks has $N_A=N_B=10^4$ and $t_r=5$. \label{gap} }
\end{figure}

We performed Monte-Carlo simulations to verify this result. First,
Fig.~\ref{gap}, shows infection densities at different $\beta$, corresponding to
a horizontal sweep across the phase diagram seen in Fig.~\ref{Phase} at $\langle
k_{AB} \rangle =0.1$. The epidemic spreading first occurs at
$\beta_c(\kappa_B)$, where the disease enters the epidemic phase and spreads
through network B, while the infection density in network A remains negligible.
This mixed region, in agreement with our predictions, occurs in the region
$\beta_c(\kappa_B) < \beta < \beta_c(\kappa_T)$. In this regime, network A plays
no role in the spreading of the infection on network B. Above
$\beta_c(\kappa_T)$, we see that the infection density in network A begins to
rise, showing that the entire interconnected network system is now in the epidemic phase, as predicted.

\begin{figure}
\includegraphics[width=6cm,height=8cm, angle=-90]{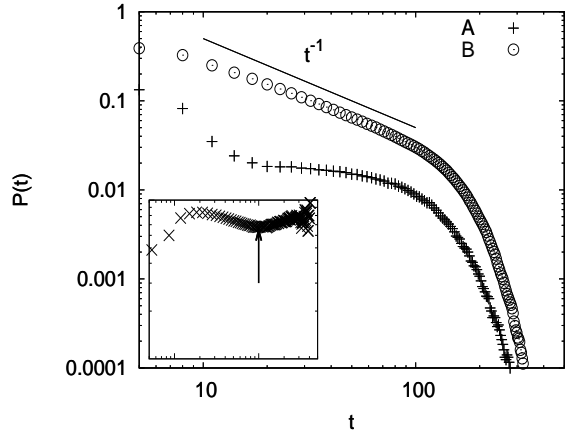}
\caption{Epidemics exhibit critical survival only in one network for weakly-coupled
 networks. Infection survival probabilities $P(t)$ on the individual networks
with internetwork connectivity $\langle k_{AB}\rangle =.1$ at $\beta =
\beta_c(\kappa_B)$. The survival probability in network A (lower curve, with +),
the less connected network, is a small fraction of that in network B (upper
curve with $\bigcirc$), the more connected network. The survival probability in network B
falls off as $t^{-1}$, as expected of a system at criticality. Network A does
not show a smooth decrease towards $0$ typical of a network much below criticality.
{The inset shows the relative difference in the survival probabilities, $[P_B(t)-P_A(t)]$/P(t). The arrow
indicates the minimum difference between the two curves, after the initial increase.}
Network parameters are $\langle k_A \rangle = 1.5, \langle k_B \rangle = 6.0$, $N_A=N_B=10^4$, and $t_r=5$. \label{P(t)4}}
\end{figure}

{ For networks approaching the strongly-coupled regime from below,
the mixed phase is expected to be small, and thus difficult to identify from graphs such as the one in Fig.~\ref{gap}.
We thus examine not only the infection densities, but also the survival probability $P(t)$, which is the probability of an infection started from a single infected site being active at a time $t$. Equivalently and more accessible from public health records, the distribution of time spans of reported outbreaks can be used.}
At criticality, the probability of an infection started from a single infected site remaining active at a later
time $t$ is expected to scale as $P(t) \sim t^{-1}$\cite{Hinrichsen}.
Fig.\ref{P(t)4}, shows the survival probabilities of the networks comprising
an interconnected with $\beta=\beta_c(\kappa_B)$ for both the strongly- and weakly-coupled cases.
In both cases, Network B exhibits the expected $t^{-1}$ fall-off
in survivability with time that is expected of a system at criticality, indicating that Network B is actually
undergoing a phase transition at the disease-free/mixed phase line.
In the weakly-coupled case, however, the survival probability in network A
does not fall off as expected, due to infrequent and non-epidemic
instances of infections from network B. The slope of the survival probability
for network A thus cannot be used directly to confirm when it enters the epidemic phase.
However,  if both networks are participating in an epidemic, the disease should be active in
both networks at each time step. {We thus introduce the survival probability gap (inset of Fig.~\ref{P(t)4}), 
$\Delta P(t) = \min\left[ \left[ P_B(t)-P_A(t)\right] /P(t)\right]$, or the minimum relative difference in the
likelihoods that each network will have any infected members (nodes in class $i$) present at time $t$.
We use this quantity to measure the deviations of the survival probability in network A from the value that
is obtained from a network at criticality (network B) and thus of how far away network A is from it's own
critical point.}

\begin{figure}
\includegraphics[width=6cm,height=8cm, angle=-90]{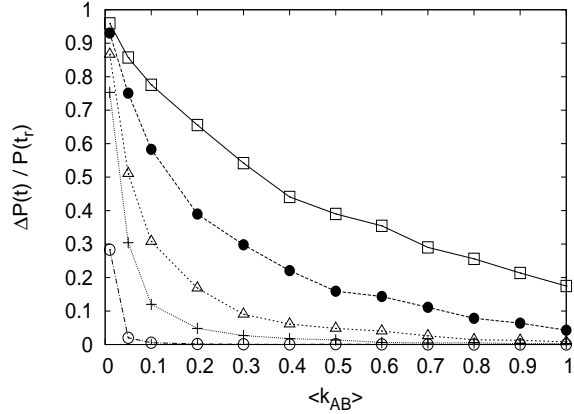}
\caption{Survival probability gap shows mixed phase boundaries. Fractional size
of the minimum survival probability gap, $\Delta P(t)/P(t_r)$, (minimum distance between the two curves in Fig.~\ref{P(t)4}  after the time $t_r$ has passed) between two interacting ER networks with $\langle k_A \rangle=1.5$ and $\langle
k_B\rangle=6.0$ at various infection strengths. From top to bottom
$\beta=\beta_c(\kappa_B)=.0358,.038,.04,.042,\beta_c(\kappa_T)=.044$. The gap
shrinks with increasing interaction and with increasing $\beta$, matching Fig.
\ref{Phase}. At $\beta_c(\kappa_T)$, the survival probability is the same in
both networks for all but $\langle k_{AB}\rangle=.01$, where it remains distinct
due to finite size effects. For all systems, $N_A=N_B=10^4$ and $t_r=5$.\label{P_gap_vanish}}
\end{figure}

In Fig.~\ref{P_gap_vanish} we plot this survival gap at different $\beta$,
equivalent to vertical slices across the phase diagram seen in
Fig.~\ref{Phase}. We see that when $\langle k_{AB}\rangle$ or $\beta$ is
increased to move outside the expected mixed phase region and into the epidemic phase, $\Delta P(t)$ goes to zero.  In other words, at the mixed/epidemic phase line, network
A is behaving identically to a network known to be at criticality, and thus can be said to itself be critical along that line.
 This confirms the assertion that there can
only be a gap in survival probability when one
network is in the epidemic phase and the other is not, i.e. in the mixed phase. 
Thus a non-zero survival probability gap can serve as a good predictor for the presence of the mixed phase.

\begin{figure}
\includegraphics[width=6cm,height=8cm, angle=-90]{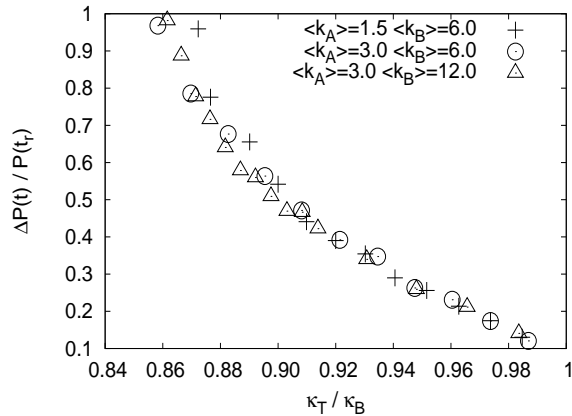}
\caption{Survival probability gap scaling is universal when rescaled by $\kappa_B$ for
different networks. The data collapses onto a
single curve, showing this universal behavior. The ratio of $\kappa_T$ to $\kappa_B$
determines $\Delta P(t)$ along the disease-free / mixed phase border. For the network systems with $\langle
k_B\rangle=6.0, \beta=.0358$, for $\langle k_B\rangle=12.0, \beta=.01725$. Both
are $\beta_c(\kappa_B)$ of the respective systems. 
Again, for all systems $N_A=N_B=10^4$ and $t_r=5$. \label{P_universal}}
\end{figure}

Lastly, we addressed the question of universality under different values
of inter- and intra-network degree, finding that along the disease-free/mixed
phase transition line, the behavior of networks with different $\kappa$ is
universal under appropriate scaling. Fig.\ref{P_universal} shows three different
networks, all with $\langle k_B\rangle > \langle k_A\rangle$ and
$\beta=\beta_c(\kappa_B)$. Rescaling the survival probabilities by $P(t_r)$ and
plotting vs. $\kappa_{T} / \kappa_{B}$ instead of $\kappa_T$ directly,
the curves collapse, showing $\Delta P(t)$, and thus the mixed phase, disappear uniformly as the network
approaches the strongly-coupled regime ($\kappa_T > \kappa_B$ and $\langle k_{AB} \rangle > \langle k_{AB} \rangle_c$.)
{Near the critical point, $\Delta P(t) \sim [(\kappa_T-\kappa_B)/\kappa_B]^{-1}$ (fit not shown.)}
This identical behavior implies survival probabilities in networks could be used as as a measure of network
connectivities near criticality, as the latter may be difficult to obtain for
social and biological networks, whereas information on the duration of an epidemic outbreak in
various communities (from which $P(t)$ can be estimated) is likely to be recorded. {In addition, the survival probability gap persists well beyond $\langle k_{AB} \rangle = 1$, for appropriate
$\langle k_A\rangle$ and $\langle k_B\rangle$. For the system with $\langle k_A \rangle = 3.0$
and $\langle k_B \rangle = 12.0$, $\langle k_{AB}\rangle_c \approx 2.47$. Even when every
node in network A is connected to two or more nodes in network B, there can still be an order of magnitude
difference at minimum in the likelihood of finding the disease active in the two networks; the mixed phase region is not confined to small $\langle k_{AB}\rangle$ only.}

\section{Conclusions}
In summary, we introduced two classes for interconnected network systems, strongly
coupled and weakly coupled, and studied the behavior of epidemics on them.
In strongly-coupled network systems, epidemics occur always across the entire interacting network system
, with the presence of interconnections enhancing
epidemic spreading. In weakly-coupled network systems, a mixed phase
exists where epidemics do not always occur across the full interconnected network system,
and interconnections affect only epidemic spreading across the less intraconnected network. We demonstrated the boundaries and behavior of the mixed
phase numerically as well as analytically. Proper analysis of which groups of communities
comprise strongly- or weakly-coupled systems could inform public
policy and highlight the necessity of cooperation between different governing
bodies or provide information about the epidemic danger of increasing interaction between human and animal populations.
\begin{acknowledgments}
We thank the DTRA, European EPIWORK project, and the ONR for financial support, SH thanks also the Israel Science Foundation.
\end{acknowledgments}

\end{document}